\documentclass[%
 reprint,
superscriptaddress,
amsmath,amssymb,
aps,
pra,
]{revtex4-2}

\usepackage{graphicx}% Include figure files
\usepackage{dcolumn}% Align table columns on decimal point
\usepackage{bm}% bold math
\usepackage{natbib}

\begin{document}

%\preprint{APS/123-QED}

\title{Induced flow inside a droplet by static electrical charge}% Force line breaks with \\

\author{Tapan Kumar Pradhan}
\email{tapan.k.pradhan@gmail.com}
 \affiliation{Department of Mechanical Engineering, \\ Indian Institute of Technology Kharagpur, Kharagpur 721302, India}%Lines break automatically or can be forced with \\
\author{T. N. Banuprasad}%
 
\affiliation{Advanced Technology Development Centre, \\ Indian Institute of Technology  Kharagpur,  Kharagpur, West Bengal 721302, India.}
\author{M. S. Giri Nandagopal}
\affiliation{Department of Mechanical Engineering, \\ Indian Institute of Technology Kharagpur, Kharagpur 721302, India}

\author{Suman Chakraborty}
\email{suman@mech.iitkgp.ernet.in}
\affiliation{Department of Mechanical Engineering, \\ Indian Institute of Technology Kharagpur, Kharagpur 721302, India}
\affiliation{Advanced Technology Development Centre, \\ Indian Institute of Technology  Kharagpur,  Kharagpur, West Bengal 721302, India.}

\begin{abstract}
Introducing controlled fluid motion inside a droplet turns out to be of outstanding scientific importance. In this work, we suggest a new method of flow manipulation inside a sessile droplet by simply deploying a static charge produced by triboelectric effect. This is physically actuated by charge transfer between the two lateral electrodes within which the droplet is entrained, triggering a strong ionized air current. The flow inside the droplet is generated due to the shear exerted at the liquid-air interface by the charged induced ionized air flow around the droplet, a paradigm that has hitherto remained unexplored. The strength of the fluid flow can be controlled by adjusting the supplied charge. Such unique controllability without sacrificing the physical simplicity opens up new possibilities of flow manipulation in a multitude of applications in droplet based microfluidics.
\end{abstract}

\maketitle

\section{Introduction}

Actuating, manipulating and controlling fluid motion in a droplet turns out to be of emerging scientific importance, as attributed to a multitude of applications ranging from engineering to biology. The fundamental premises that control their functionalities include efficient mixing on one side and arresting unwarranted deposition of particulate matters on the other\cite{Deegan1997,Mampallil2011,Eral2011,Fischer2002}. In evaporating droplets, the flow mainly occurs by capillary action \cite{Deegan1997}, surface tension gradient \cite{Hegseth1996,Karpitschka2012,Semenov2017,Karpitschka2017,Darras2018,Marin2019} or buoyancy forces \cite{Kang2013,Pradhan2016,Edwards2018,Li2019}. However, because of their limited controllabilities, several external actuation mechanisms have been employed to generate desired flow characteristics within a droplet. In recent years, electrical forcing has emerged to be a preferentially alternative mechanism for flow manipulation inside tiny droplets, primarily due to facilitated integrability of a droplet with electrical or electronic circuitry from the emerging perspective of lab on a chip technology \cite{Taylor1966,Kunti2018,Lee2009,Chang2007,Mugele2006}. Despite their obvious merits, reported techniques on electrically driven flow manipulation inside droplets suffer from a number of compelling deficits. In particular, these processes inevitably demand specialized fabrication procedures for the electrodes and surface modification. In addition, it often necessitates changing the composition of fluid to modify the ionic strength, losing the freedom of being applicable to the desired applications on-demand.

\par
Overcoming the above constraints, we report here an approach of flow manipulation inside a sessile droplet by simply harnessing a static charge produced by triboelectric effect. We realize this by charge transfer between the two lateral electrodes within which the droplet is entrained, without demanding any on-chip electrode fabrication. Fluid motion inside the droplet is generated due to the interfacial shear exerted by the charged induced ionized air flow incipient to the droplet. The strength of the resultant fluid motion can be controlled at will, by altering the supplied charge. Simplicity of the technique, furthered by its elegant controllability, renders this ideally suitable for a wide gamut of applications ranging from droplet microreactors to digital microfluidics. 

\begin{figure}[htb!]
\begin{center}
\includegraphics[width=0.48\textwidth]{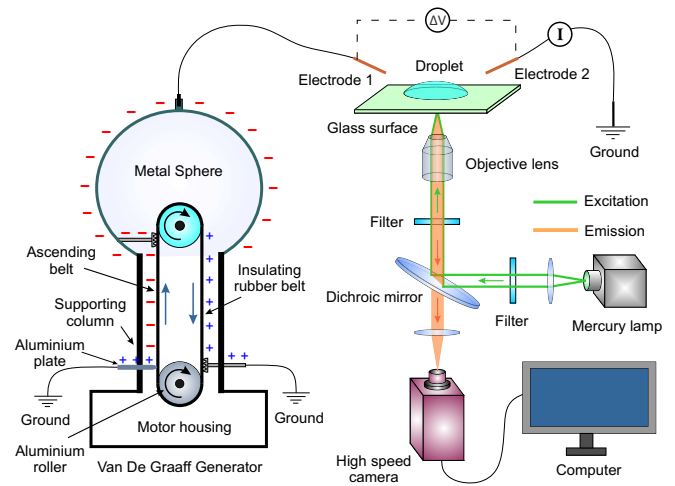}% Here is how to import EPS art
\caption{\label{fig:exp} Experimental setup for inducing flow inside the droplet and micro-PIV measurement.}
\end{center}
\end{figure}

\section{Experimental Method}
A water (DI) droplet of volume ($\forall$) 0.7 $\mu$L was placed on a clean glass surface forming a contact angle of 36 $^0$ and contact line diameter of 2.1 mm. The droplet was placed between two copper electrodes of 0.5 mm diameter having separation distance of 8 mm. The electrodes were physically separated from the droplet with air gap on both the sides (Fig. \ref{fig:exp}). Static charge generated by a Van De Graff generator (VDG) was applied to induce flow inside the droplet.  One of the electrodes (electrode 1) was connected to the VDG and the other electrode (electrode 2) was connected to the ground. VDG works on the principle of triboelectric effect where friction between two dissimilar materials creates unbalanced charge on each material. In the present setup (Fig. \ref{fig:exp}), friction between the rubber belt and the aluminium plate (present near the lower roller) generates negative charge on the belt. The negative charge from the belt was transferred to the dome which then transferred to the droplet by a connecting wire. The electrons from the VDG flow to the ground through the electrode 1, droplet and electrode 2. The current ($I$) in the circuit and the voltage between the two electrodes ($\Delta V$) were measured by two multimeters as shown in Fig. \ref{fig:exp}. The effect of charge on the flow was carried out by regulating the current flow over a range of 1 $\mu$A to 4.7 $\mu$A by adjusting the roller speed.

\par
The velocity of fluid inside the droplet was measured by micro-particle image velocimetry (PIV) technique. Fluorescent polystyrene particles of diameter 4 $\mu$m were added to the droplet as tracer particles which were illuminated by a mercury lamp. High speed camera (Phantom V15) attached with an inverted microscope (Olympus NX71) was used to capture the images at 200 frames per second (fps). Each image has a field of view of 2.65 mm $\times$ 2.65 mm. The images were processed by PIVLab \cite{Thielicke2014} to get velocity field using FFT algorithm with 64 $\times$ 64 interrogation area. 

\begin{figure}[htb!]
\begin{center}
\includegraphics[width=0.47\textwidth]{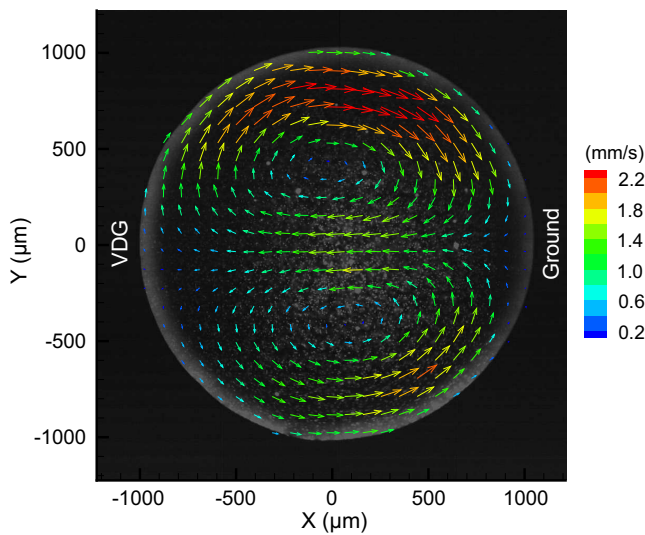}% Here is how to import EPS art
\caption{\label{fig:vel} Velocity field inside the droplet near the substrate at Z= 50 $\mu$m from the substrate surface. The measured current in the circuit is equal to 2 $\mu$A.}
\end{center}
\end{figure}

\section{Results and Discussion}
When no charge is supplied to the electrode, the internal convection shows very small flow strength of the order of 1 $\mu$m/s \cite{Pradhan2015}. The fluid moves unidirectionally towards the contact line which is caused by capillary action to compensate the higher evaporative flux at the contact line \cite{Deegan1997}. When the charge is applied to the electrode (electrode 1 in Fig. \ref{fig:exp}) connected to the VDG, the fluid inside the droplet shows strong convection as shown in Fig. \ref{fig:vel}. It shows two circulating loops (supplementary movie A \cite{movie}) having maximum flow strength of the order of 2.5 mm/s. No flow is observed when the circuit is disconnected from the ground. Flow is only observed when the charge continuously flows from the VDG to the ground through the droplet.

\begin{figure}[htb!]
\begin{center}
\includegraphics[width=0.47\textwidth]{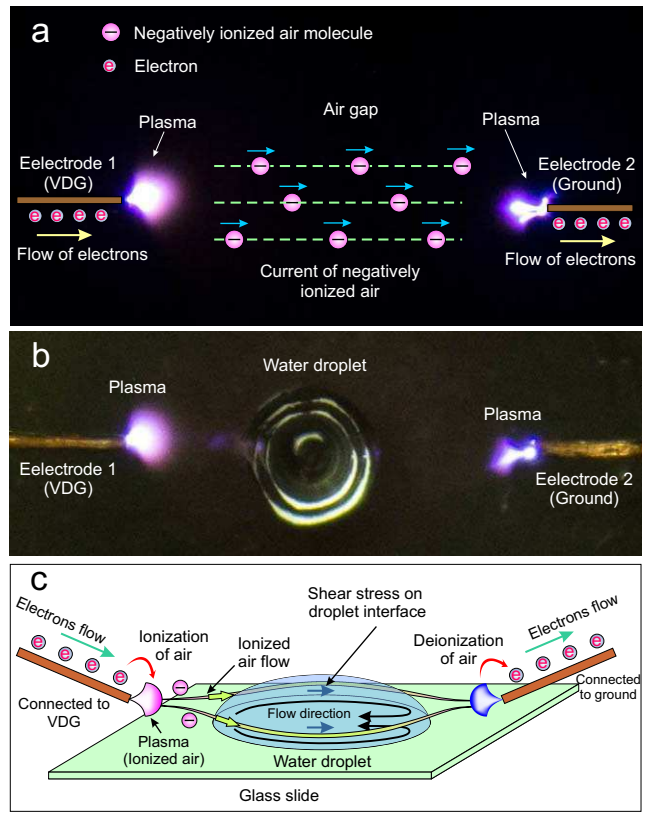}% Here is how to import EPS art
\caption{\label{fig:mechanism} Plasma formation near the electrodes and flow mechanism. (a) Depiction of ionized air molecules between the two electrodes. (b) Droplet placed between the two electrodes with charge transfer between the electrodes. (c) Depiction of the mechanism of induced flow.}
\end{center}
\end{figure}

\par
The negative charge from the VDG is carried to the electrode 1 by a connecting wire. Electrode 1 carries unbalanced electrons and tries to get neutralized. Electrode 2 which is connected to the ground is the easiest path to get neutralized. Hence, the electrons tend to move from electrode 1 to electrode 2. During the process, collision between the electrons and the air molecules surrounding the electrode 1 creates plasma of negatively charged air. Air molecules acquire negative charge due to electron capture ionization when the molecules are subjected to electron impact \cite{Relkin1987,Hiraoka1977}. The plasma formation near the electrodes is presented in Fig. \ref{fig:mechanism} (a),(b). The image of the plasma was captured in a closed dark chamber using a DSLR camera keeping high exposure time (15 sec). The intensity of the plasma is very less due to low current and the tiny plasma can only be visible in a closed dark chamber. The negatively charged air molecules near electrode 1 accelerate towards the ground electrode (electrode 2) to get neutralized which is depicted in Fig. \ref{fig:mechanism} (a). It creates a flow of negatively charged air from electrode 1 to electrode 2, maintaining a continuity of charge flow in the circuit. The static electricity works different than the normally used DC and AC supplies. In such a low voltage (below the breakdown voltage) of normally used DC and AC, the charge cannot pass between the two electrodes having an air gap of 8 mm. In case of static electricity, the charged molecules can be transferred between these electrodes due to a small potential difference similar to electron acceleration in cathode ray tube. Also the observed plasma is not a continuous electric spark which is observed at high DC and AC breakdown voltage. Too close separation distance between the electrodes causes spark and larger separation distance stops the flow of charge between the electrodes. An approximate intermediate separation distance is choosen to maintain the charge flow without spark.

\par
The schematic of the induced flow inside the droplet using static charge is presented in Fig. \ref{fig:mechanism}(c). When the droplet is placed between the two electrodes as shown in Fig. \ref{fig:mechanism}(b), the droplet experiences the air flow between the electrodes. The flow of air between the two electrodes is confirmed by the smoke flow visualization using an incense stick (supplementary movie B \cite{movie}). Smoke from the incense stick drifted in the direction of air flow between the two electrodes. The smoke shows an air flow from the electrode connected to the VDG to the ground electrode. The surrounding air current exerts a shear stress at the liquid-air interface of the droplet. The shear stress at the interface induces circulating flow inside the droplet \cite{Horton1965} which is well explained in the Fig. \ref{fig:mechanism}(c). The flow pattern inside the droplet strongly depends on the air flow pattern near the droplet. Any disturbance in the surrounding air significantly affects the internal convection of the droplet.

\par
The water droplet evaporates at ambient temperature ($T_{\infty}$) of $25 ^0$C and relative humidity ($\phi$) of 0.6. The evaporation time ($t_f$) of the sessile droplet for small contact angle without any applied charge is given by \cite{Brutin}:

\begin{equation}
  t_f \approx \frac{\pi \rho R^2 \theta _ 0}{16 D_v (C_s - C_{\infty})} 
\end{equation}

\par
The droplet has a constant contact line radius ($R$) throughout the evaporation process which is equal to 1.05 mm. The value of the coefficient of diffusivity ($D_v$), saturated vapor concentration ($C_s$), ambient vapor concentration (\( C_{\infty} = \phi C_s\)), and density of water ($\rho$) are equal to \(2.42 \times 10^{-5} \; \textrm{m}^2/\textrm{s}\) \cite{crc}, \(2.3 \times 10^{-2} \; \textrm{kg/m}^3\) \cite{crc}, \(1.38 \times 10^{-2} \; \textrm{kg/m}^3\), and 997 $kg/m^3$ \cite{crc} respectively. The evaporation time ($t_f$) of the water droplet calculated from the above equation is equal to 10.2 min without any applied charge. When charge is applied, it may add thermal energy due to joule heating or plasma to the droplet affecting evaporation dynamics and internal flow. Maximum possible thermal energy absorbed by the droplet from the applied charge is equal to the total electrical energy lost in the process, which is given by, \( W = \Delta V I t_e\). Here, $t_e$ is the evaporation time of the droplet when the source of energy for evaporation comes only from the applied static charge. The thermal energy required for complete evaporation of the droplet is given by, \(H=m h_{fg}\). Here, $m$ is the mass of the water droplet (\(m = \rho \forall \)) and $h_{fg}$ is the enthalpy of vaporization at 25 $^0$C. The value of $\Delta V$ and $I$ are equal to 32 V and 2 $\mu$A respectively for the velocity field shown in Fig. \ref{fig:vel}. Equating the thermal energy for vaporization and supplied electrical energy:

\begin{equation}
  t_e = \frac{m h_{fg}}{\Delta VI}
\end{equation}

\par
The value of ($t_e$) is equal to 443 min. The evaporation time of the droplet due to only electrical power from VDG is much more than the normal evaporation time of the droplet evaporating in room environment by extracting energy from the substrate. Hence, the effect of the heat of plasma or electrical energy on evaporation of the droplet is neglected.  VDG has disadvantages for many applications due to its low current. However, low current is advantageous in the present application, because it leads to low total energy, keeping the droplet unaffected. Hence, it avoids rapid evaporation of the droplet. Due to very low thermal energy, the thermal effect like thermal Marangoni convection on the fluid flow inside the droplet can be neglected. 

\par
Other possible flow mechanisms caused by electrical energy may contribute to the observed flow are electrohydrodynamics, electrothermal, electro-osmosis and electrophoresis. EHD causes quadratic flow structure inside the droplet, \cite{Taylor1966} whereas only two circulating loops (Fig. \ref{fig:vel}) are observed in the present experiment.  Hence, it is expected that the flow is not due to the EHD effect. Electrothermal effect is due to electric body force caused by Joule heating. However, from the above time scale analysis of evaporation, it is found that the supplied electrical energy hardly causes any thermal effect. An approximate velocity magnitude for electro-osmotic flow can be given by the Helmholtz-Smoluchowski equation, \(u_{eof} = \frac{\epsilon \zeta E}{\mu}\). The value of electric permittivity of water ($\epsilon $), viscosity of water ($\mu$), and zeta potential of water-glass interface ($\zeta$) are equal to \(6.96 \times 10^{-10} \, F/m\) \cite{crc}, \(8.9 \times 10^{-4}\) Pa.s \cite{crc}, and  -62.2 mV \cite{Gu2000} respectively. The voltage across the electrodes is 32 V which gives average electric field strength ($E$) of \(4 \times 10^3\) V/m.  The velocity scale is equal to 0.2 mm/s which is negligible as compared to the observed value. Polystyrene particles (diameter = 4 $\mu$m) are used in the fluid for micro-PIV measurements which may show electrophoretic movement in the electric field. Considering the particles much more larger than the EDL, the electrophoretic velocity can be given by, \(u_{ep}=  \frac{\epsilon \zeta _p E}{\mu}\). The approximate value of zeta potential of the polystyrene particle-water ($\zeta_p$) is equal to -79 mV \cite{Ho1997}. The electrophoretic velocity scale is equal to 0.25 mm/s which is very less compared to the observed velocity. In the present experimental setup, the electrons show a unidirectional flow from VDG to the ground which can be resembled as DC. The DC electro-osmotic flow and electrophoretic movement of the particles are expected to show a unidirectional flow \cite{Kim2006} rather than a circulating flow as observed in the present case. Hence, the electro-osmotic flow and electrophoretic movement are considered to be absent. The observed flow pattern in the present situation, therefore, can only be explained by the shear stress induced flow due to the external ionized air flow.

\begin{figure}[htb!]
\begin{center}
\includegraphics[width=0.45\textwidth]{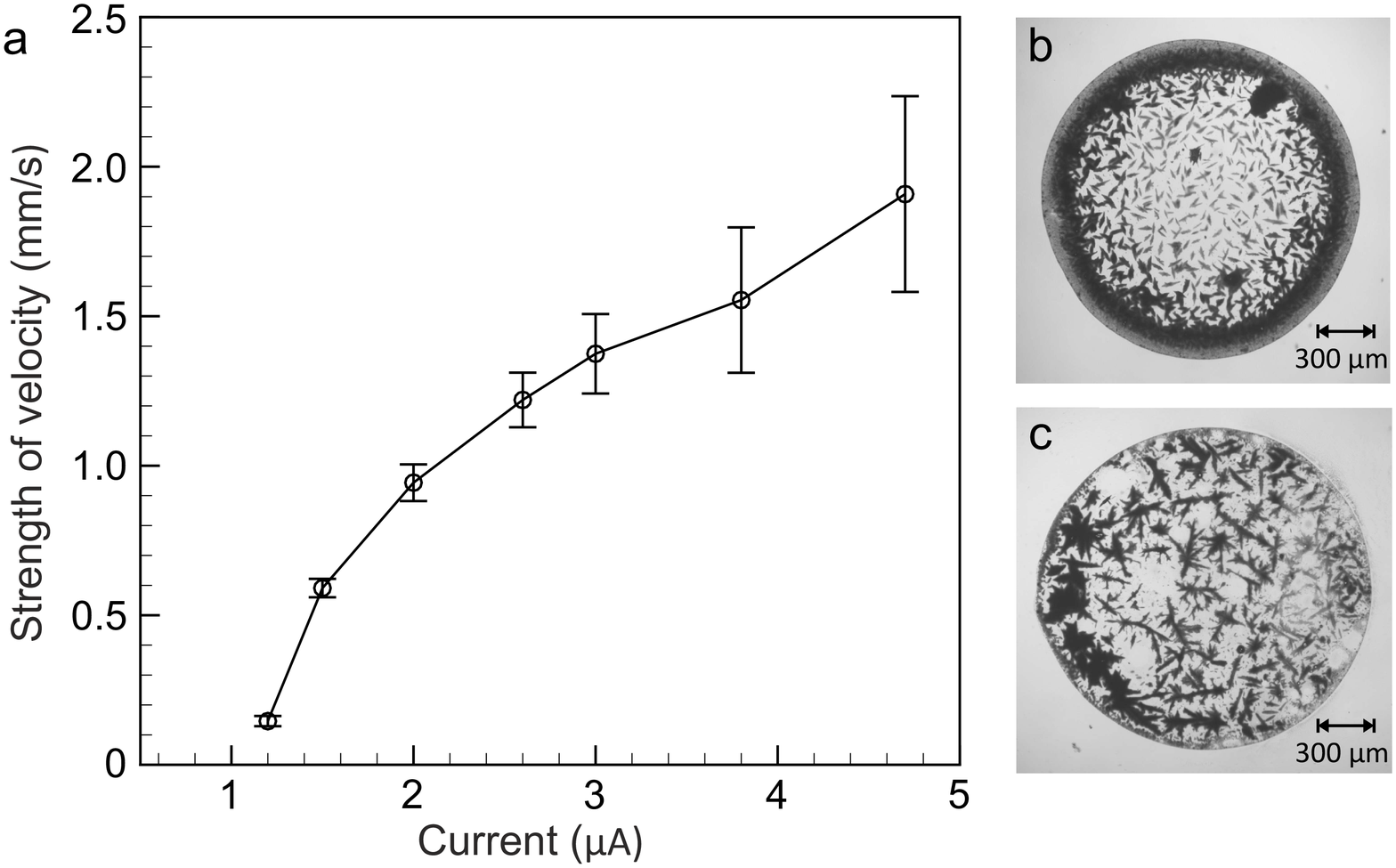}% Here is how to import EPS art
\caption{\label{fig:strengthdep} (a) Variation of flow strength inside the droplet with charge flow. (b) Deposition pattern of drying droplet of fountain pen ink without applied charge. (c) Deposition pattern of drying droplet of fountain pen ink with applied charge.}
\end{center}
\end{figure}

\par
We next study the effect of charge on the flow strength at different charge flows which is presented in Fig. \ref{fig:strengthdep}(a). The flow strength is characterized by area average of velocity magnitude. With increase in charge flow in the circuit, the negatively ionized air molecules increase in number. This leads to increase in flow rate of air surrounding the droplet. Hence, the flow strength inside the droplet increases with increase in charge flow between the two electrodes. It is also observed from the plot that the uncertainty (error bar) increases with increase in current flow. It is expected that the increase in air flow rate increases the fluctuation in the air velocity and becomes more chaotic.

\par
The flow induced by the static electricity can also be used as a method to manipulate the deposition pattern of a drying droplet. Deposition pattern of a drying droplet strongly depends on the flow pattern inside the droplet. The coffee ring deposit of a droplet of fountain pen ink is shown in Fig. \ref{fig:strengthdep}(b) which occurs due to the unidirectional capillary flow caused by evaporation \cite{Deegan1997}. When the droplet is dried with simultaneous application of static charge to the droplet, the coffee ring breaks and forms a deposition throughout the droplet base area (Fig. \ref{fig:strengthdep}(c)). The breaking of coffee ring occurs due to the flow circulation (Fig. \ref{fig:exp}(b)) caused by the applied charge. The circulation suppresses the unidirectional capillary flow and disturbs the particle convection towards the contact line. Though electric field has been applied to observe different phenomena in droplet \cite{Taylor1966,mchale2011,Hamlin2012,Brandenbourger2016}, use of static charge in fluid manipulation has not been explored yet. Present work may open many opportunity to couple static electricity with fluid mechanics and droplet dynamics.

\section{Conclusion}

In summary, we have suggested a simple method of inducing controlled flow inside a droplet, mediated by static electric charges. The charge is produced by triboelectric effect using a Van De Graff generator. The fluid motion inside the droplet is generated by the shear stress exerted at the liquid-air interface by the charge induced air flow around the droplet. This imposes no fabrication constraints on the fluidic system, and puts no limitation on the nature of the fluid. Also, there is no change in the composition of fluid during the process. Biological sample can also be used in the droplet as there is no heating and composition change during the flow manipulation.

\begin{acknowledgements}

SC acknowledges Department of Science and Technology, Government of India, for Sir J. C. Bose National Fellowship. MSGN acknowledges DST-SERB for the National Post-Doctoral Fellowship (Sanction No: PDF/2018/001486). All the authors acknowledge Indian Institute of Technology Kharagpur for financial support.
\end{acknowledgements}

\nocite{*}
\bibliography{reference}% Produces the bibliography via BibTeX.

\end{document}